% ------------------------------------------------------
%    eipre.tex   July 1998   (preprint version Sept 98)
% ------------------------------------------------------
\def\ptitle{An inversion inequality for potentials in QM}
% ------------------------------------------------------------------------
% Change list publication -> preprint
% (1) Remove magnification at start
% (2) Remove \parskip and \baselineskip lines before Abstract
% (3) Remove \np before and after references
% (4) Add \input psfig.sty (or some such) at top
%     Add \psfig commands for each figure
% ------------------------------------------------------------------------
\nopagenumbers
%\magnification=\magstep1
\hsize 6.0 true in 
\hoffset 0.25 true in 
% 6 in width with 1.25 in margins default = (6.5, 0)
\emergencystretch=0.6 in                 % TEXBook p 107 : allows h-space 
\vfuzz 0.4 in                            % page-length flexibility
\hfuzz  0.4 in                           % line-length flexibility
\vglue 0.1true in
\mathsurround=2pt                        % Default is 2pt
\topskip=24pt                            % Default is 10pt
                     % New line
\def\np{\hfil\vfil\break}                % New page
\def\title#1{\bigskip\noindent\bf #1 ~ \tr\smallskip} % Headings
\baselineskip = 16 true pt
% --------------------------------------------------------------------
%   PC Fonts (used only locally)
% --------------------------------------------------------------------
%\font\tr=TIMENRR                       % Times New Roman: see output
%\font\bf=TIMENRB                       % Redefinition
%\font\it=TIMENRRI                      % Redefinition       
%\font\trbig=TIMENRR scaled \magstep3   % For main Title
%\font\th=CMBXSL10                      % Theorems                    
%\font\tiny=CMBX8                       % Page numbers
% --------------------------------------------------------------------
%  generic unix fonts (lower case names)
% --------------------------------------------------------------------
\font\tr=cmr12                          % Our default
\font\bf=cmbx12                         % Redefinition
\font\sl=cmsl12                         % Redefinition
\font\it=cmti12                         % Redefinition
\font\trbig=cmbx12 scaled 1500          % Main Title
                          % Theorems                       
\font\tiny=cmr10                        % Running title
% --------------------------------------------------------------------
             % Math Sets eg R -> |R
                 % bold in math mode
                % small bold in math mode
\def\ng{>\kern -9pt|\kern 9pt}          % not greater than
\def\bra{{\rm <}}                       % bra  <  math mode
\def\ket{{\rm >}}                       % ket  >  math mode
\def\hi#1#2{$#1$\kern -2pt-#2}          % hyphen \hi{N}{body} = N-body
\def\hy#1#2{#1-\kern -2pt$#2$}          % hyphen hy{large}{N} = large-N

% --------------------------------------------------------------------- boxes ---
\def\dbox#1{\hbox{\vrule  %  Open box size 2#1 (Abrahams p 273) 
                        \vbox{\hrule \vskip #1
                             \hbox{\hskip #1
                                 \vbox{\hsize=#1}%
                              \hskip #1}%
                         \vskip #1 \hrule}%
                      \vrule}}

\def\qed{\hfill \dbox{0.05true in}}  %  QED
 % SQUARE 
% ---------------------------------------------------------------------- output -------
\output={\shipout\vbox{\makeheadline
                                      \ifnum\the\pageno>1 {\hrule}  \fi 
                                      {\pagebody}   
                                      \makefootline}
                   \advancepageno}

\headline{\noindent {\ifnum\the\pageno>1 
                                   {\tiny \ptitle\hfil page~\the\pageno}\fi}}
\footline{}
% ---------------------------------------------------------------- begin our ref.tex --
\newcount\zz  \zz=0  % switch for printing references
\newcount\q   %  reference number
\newcount\qq    \qq=0  % starting reference number-1   (usually zero)

\def\pref #1#2#3#4#5{\frenchspacing \global \advance \q by 1     % paper reference
    \edef#1{\the\q}
       {\ifnum \zz=1 { %
         \item{[\the\q]} 
         {#2} {\bf #3},{ #4.}{~#5}\medskip} \fi}}

\def\bref #1#2#3#4#5{\frenchspacing \global \advance \q by 1     % book reference
    \edef#1{\the\q}
    {\ifnum \zz=1 { %
       \item{[\the\q]} 
       {#2}, {\it #3} {(#4).}{~#5}\medskip} \fi}}

\def\gref #1#2{\frenchspacing \global \advance \q by 1  % general reference
    \edef#1{\the\q}
    {\ifnum \zz=1 { %
       \item{[\the\q]} 
       {#2}\medskip} \fi}}

 \def\sref #1{~[#1]}

\def\references#1{\zz=#1
   \parskip=2pt plus 1pt   % default is 0pt plus 1pt       
   {\ifnum \zz=1 {\noindent \bf References \medskip} \fi} \q=\qq

  \bref{\chad}{K. Chadan and P. C. Sabatier}{Inverse Problems in
Quantum Scattering Theory}{Springer, New York, 1989}
{The `inverse problem in the coupling constant' is discussed on p406}
\pref{\halla}{R. L. Hall, Phys. Rev A}{51}{1787 (1995)}{}          % WKB
\pref{\hallb}{R. L. Hall, J. Phys. A:Math. Gen}{25}{4459 (1992)}{} % refining theorem
\pref{\hallc}{R. L. Hall, Phys. Rev. A}{50}{2876 (1995)}{}         % flat bottoms
\pref{\halld}{R. L. Hall, J. Phys. A:Math. Gen}{28}{1771 (1995)}{} % geom spect inversion
\pref{\halle}{R. L. Hall, J. Math. Phys.}{25}{2708 (1984)}{}       % kinetic potentials
\bref{\fel}{W. Feller}{An Introduction to Probability Theory and its Applications}
{J.Wiley, New York, 1971}{Jensen's inequality is discussed on p 153.}
\bref{\gel}{I. M. Gelfand and S. V. Fomin}{Calculus of Variations}
{Prentice-Hall, Englewood Cliffs, 1963}{Legendre transformations are discussed on p 72.} 
 }

 \references{0}    % Initialization of reference numbers
% ------------------------------------------------------------ end our ref.tex -----

% ------------------------------------------------------ 
%   Title page and Abstract
% ------------------------------------------------------
\vskip 1.0true in
\centerline{\trbig An Inversion Inequality for Potentials}
\vskip 0.1true in
\centerline{\trbig in Quantum Mechanics}
\vskip 0.5true in
\tr % our standard font
\baselineskip 12 true pt % for address only                                     
\centerline{\bf Richard L. Hall}\medskip
\centerline{\sl Department of Mathematics and Statistics,}
\centerline{\sl Concordia University,}
\centerline{\sl 1455 de Maisonneuve Boulevard West,}
\centerline{\sl Montr\'eal, Qu\'ebec, Canada H3G 1M8.}
\centerline{email:\sl~~rhall@cicma.concordia.ca}
\bigskip\bigskip

\baselineskip 16 true pt

% ---------------------------------------------------------------------
% spacing: remove next two lines for preprint version
%\parskip=5pt plus 1pt             % MAIN PARSKIP
%\baselineskip = 22true pt         % baselineskip
% ---------------------------------------------------------------------
\centerline{\bf Abstract}\medskip
% ---------------------------------------------------------------------
We suppose: (1) that the ground-state eigenvalue $E = F(v)$ of the Schr\"odinger
 Hamiltonian $H = -\Delta + vf(x)$ in one dimension is known for all values of the
 coupling $v > 0;$  and (2) that the potential shape can be expressed in the 
form $f(x) = g(x^2),$ where $g$ is monotone increasing and convex.  The inversion
 inequality $f(x) \leq \bar{f}({1 \over {4x^2}})$ is established, in which the
 `kinetic potential' $\bar{f}(s)$ is related to the energy function $F(v)$ by the
 transformation: $\{\bar{f}(s) = F'(v),\quad s = F(v) - vF'(v)\}.$  As an example
 $f$ is approximately reconstructed from the energy function $F$ for the potential
 $f(x) = ax^2 + b/(c+x^2).$ 
\medskip\noindent PACS~~03 65 Ge

\np
% ------------------------------------------------------ 
  \title{1.~~Introduction}
% ------------------------------------------------------
\input psfig.sty

We suppose that a discrete eigenvalue $E = F(v)$ of the Schr\"odinger Hamiltonian
$$H = -\Delta + vf(x)\eqno{(1.1)}$$
is known for all sufficiently large values of the coupling parameter $v > 0$ and we
 try to use this data to reconstruct the potential shape $f.$  The usual `forward'
 problem would be: given the potential (shape) $f,$ find the energy trajectory $F;$
 `geometric spectral inversion' is the inverse of this, that is to say
 $F \rightarrow f.$

This problem should be distinguished from the `inverse problem in the coupling
 constant' discussed, for example, by Chadan and Sabatier\sref{\chad}. In this
 latter problem, the discrete part of the `input data' is a set $\{v_{i}\}$ of
 values of the coupling constant that all yield the identical energy eigenvalue
 $E.$ The index $i$ might typically represent the number of nodes in the
 corresponding eigenfunction.  In contrast, for the problem discussed in the
 present paper, $i$ is kept fixed and the input data is the graph $(F_{i}(v),v),$
 where the coupling parameter has any value $v > v_c,$ and $v_c$ is the critical
 value of $v$ for the support of a discrete eigenvalue with $i$ nodes.   There are 
strong indications on the basis of studies involving the inversion of the WKB
 approximation\sref{\halla} that inversion with a fixed $i$ becomes more efficient
 as $i$ increases (and the problem becomes more classical).  However, the present
 paper will be concerned only with inversion from the ground-state energy function
 $F_{0}(v) = F(v).$

By making suitable assumptions concerning the class of potential shapes, theoretical
 progress has already been made with this inversion problem\sref{\hallb-\halld}.
 In Ref.\sref{\hallc} a  `concentration lemma' is proved.  If we suppose that
 $H\psi = E\psi$ and $||\psi|| = 1,$ this lemma quantifies the monotone increase 
in concentration towards $x = 0$ of the probability density $\psi^{2}(x,v)$ with
 increasing $v.$   In Ref.\sref{\halld} this lemma is used to establish the
 uniqueness of the potential shape $f$ corresponding to a given energy function 
$F.$  The class of potentials for which this uniqueness proof applies are those
 non-constant potential shapes $f$ which are symmetric, continuous at $x = 0,$ 
piecewise analytic, and monotone increasing for $x > 0.$  The `envelope inversion'
 discussed in Ref.\sref{\halld} involved a class of potentials that could be expressed
 as a smooth monotone transformation $f(x) = g(h(x))$ of a soluble potential $h(x).$
 The approximation obtained was {\it ad hoc} in the sense that nothing was known
 {\it a priori} concerning the relationship between the approximation and the (unknown)
 exact potential corresponding to the given energy function $F(v).$  In the present
 paper we establish an inversion {\it inequality} for a special case of envelope
 inversion, namely the case in which the `envelope basis' is the harmonic-oscillator
 shape $h(x) = x^2.$   Thus we assume that the potential shape $f(x)$ has the
 representation 
$$f(x) = g(x^2),\eqno{(1.2)}$$
where $g$ is monotone increasing and convex ($g'' > 0$).  This is a strong
 assumption but, as we prove in Section (2),  it yields a corresponding
 strong result, that is to say:
$$f(x) \leq \bar{f}\left({1 \over {4x^2}}\right ),\eqno{(1.3)}$$
where $\bar{f}(s)$ is the `kinetic potential' corresponding to the potential
 $f(x).$  The parameter $s$ is equal to the mean kinetic energy $\bra -\Delta\ket$
 and, in terms of $s,$  the eigenvalue $F(v)$ may be represented\sref{\halle}
 {\it exactly} by the semi-classical expression:

$$E = F(v) = \min_{s > 0}\left\{s \ + v\bar{f}(s)\right\}.\eqno{(1.4)}$$

The transformations $F \leftrightarrow \bar{f}$ are essentially Legendre
 transformations\sref{\gel}.  This is so because we know\sref{\hallc} that
 $F$ and $\bar{f}$ have definite and opposite convexity; more particularly,
 we know
$$\bar{f}''(s) F''(v) = -{1 \over {v^3}} < 0.\eqno{(1.5)}$$
The transformation in the direction needed here $F \rightarrow \bar{f}$ will
 be given explicitly in Section~(2) below where we also prove the inequality
 (1.3), the main result of this paper.  In Section~(3) we discuss an example
 for which we compare the upper approximation given by (1.3) with the corresponding
 exact result.

% ------------------------------------------------------
    \title{2.~~Proof of the inversion inequality}
% ------------------------------------------------------
We suppose that the exact normalized wave function corresponding to the potential
 $vf(x)$ is given by $\psi(x,v),$ where the coupling parameter $v > 0.$ 
 Thus $(\psi,H\psi) = F(v).$  We know how this total expectation value is
 divided between kinetic and potential energies for, in more detail, we have

$$
\eqalign{
&\bra -\Delta\ket = (\psi,-\Delta\psi) = F(v) - vF'(v) = s,\cr
&\bra f\ket = (\psi,f\psi) = F'(v) = \bar{f}(s).}\eqno{(2.1)}$$

\noindent These equations also define the kinetic potential $\bar{f}(s)$
 parametrically in terms of the parameter $v > 0.$  We first use Heisenberg's
 uncertainty inequality which gives us

$$\bra -\Delta\ket \bra x^2\ket = s \bra x^2\ket\ \geq \ {1 \over 4}.\eqno{(2.2)}$$

\noindent We now consider
 
$$\bar{f}(s) = \bra f(x)\ket = \bra g(x^2)\ket \ \geq  \ g(\bra x^2 \ket).
\eqno{(2.3)}$$

\noindent This inequality follows from Jensen's inequality\sref{\fel} and the
 fact that $g$ is convex.  By applying (2.2) in (2.3) we find

$$\bar{f}(s) \ \geq g\left({1 \over {4s}}\right )=
 f\left({1 \over {2\sqrt{s}}}\right ).\eqno{(2.4)}$$

\noindent Finally by letting $x = {1 \over {2\sqrt{s}}}$ we establish the
 inversion inequality

$$f(x) \ \leq \ \bar{f}\left({1 \over {4x^2}}\right ).\eqno{(2.5)}$$
\hfil\qed

Since the transformation in the direction $F \rightarrow \bar{f}$ is already
 expressed by (2.1), the upper approximation provided by the inversion inequality
 is now completely determined.

% ------------------------------------------------------
    \title{3.~~An example}
% ------------------------------------------------------
We consider the potential shape given by

$$f(x) = ax^2 + b/(c+x^2),\quad a,b,c > 0.\eqno{(3.1)}$$

The case $a = b = c = 1$ is illustrated in Fig.(1) which shows the potential shape
 $f(x),$ in the inset graph, and also the ground-state energy function $F(v)$
 generated from it.  In Fig.(2) the upper approximation $A$ obtained by the
 inversion inequality is shown along with the exact potential shape $f$ itself.
  The set of corresponding `exact' wave functions $\psi(x,v)$ are also shown for
 $3\times 10^{-4} \leq v \leq 10.$   The wave-function normalization is
 arbitrarily taken here to be $\psi(0,v) = 20,$ so that the graphs fit on 
the same figure as the potentials.  As the coupling $v$ increases, the wave
 functions become monotonically more concentrated near zero, in agreement
 with the `concentration lemma' mentioned in Section~(1).   

% ------------------------------------------------------
      \title{3.~~Conclusion}
% ------------------------------------------------------
Although the assumption behind the inversion inequality is strong, the fact that
 such an inequality exists may be important, especially if it can eventually be
 generalized.  The expression of this result in terms of kinetic potentials could
 be avoided in principle.  However, the representation of the energy functions
 $F(v)$ in terms of $\bar{f}(s)$ has already yielded some very effective bounds
 in the forward direction and it is natural to explore this same apparatus for
 the more difficult inversion problem.  For example, in the forward direction 
the envelope method \sref{\halle} may be expressed succinctly as
$$f(x) = g(h(x))\quad\Rightarrow\quad \bar{f}(s) \approx g(\bar{h}(s)),\eqno{(3.1)}$$
where $\approx = \geq$ if $g$ is convex and $\approx = \leq$ if $g$ is concave.
   Once one has such an approximation for $\bar{f}(s),$ it can immediately be
 inserted in the expression (1.4) to yield an approximation for the corresponding
 eigenvalue $E = F(v).$ In the present paper we have found one case $h(x) = x^2$
 for which an inequality is retained for the inverse problem.  In Ref.\sref{\halld}
 we also explored the idea of inverting the Rayleigh-Ritz variational method and
 we obtained an inversion approximation with respect to a chosen family of `trial'
 functions.  However, unlike the situation in the forward direction, the inversion
 approximation obtained was again {\it not} an inequality.   Our experience with
 this problem so far suggests that it is difficult to generate potential
 inequalities for geometric spectral inversion.   

% ------------------------------------------------------   
   \title{Acknowledgment}
% ------------------------------------------------------
Partial financial support of this work under Grant No. GP3438 from the Natural
 Sciences and Engineering Research Council of Canada is gratefully acknowledged.
% ------------------------------------------------------ 
  \references{1}
% ------------------------------------------------------
% \np

\hbox{\vbox{\psfig{figure=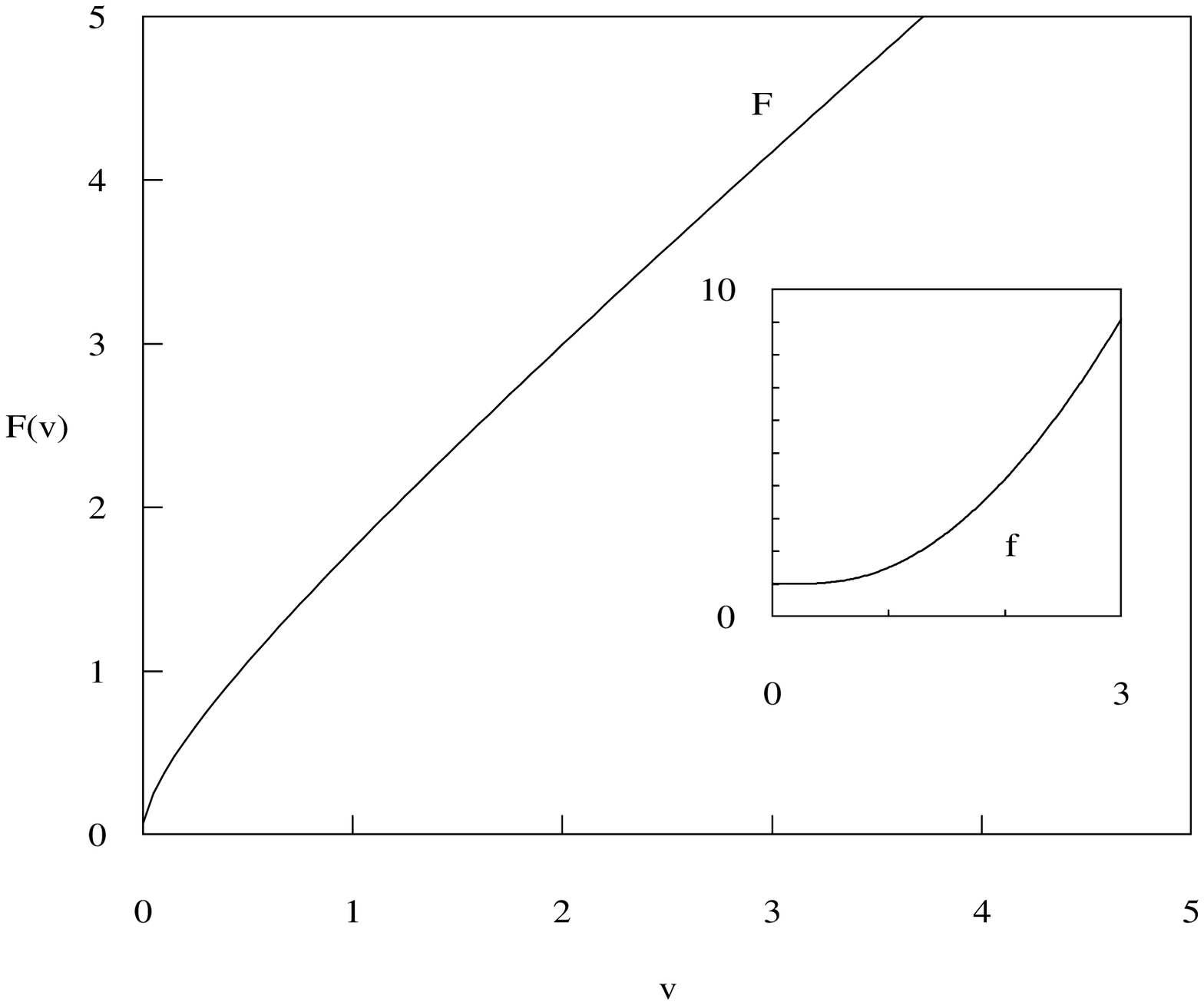,height=6in,width=6in,silent=}}}
\noindent {\bf Figure~(1)}~~The potential $f(x) = x^2 + 1/(1+x^2)$ is shown
 in the inset graph, along with the corresponding ground-state energy function
 $E = F(v).$  The aim of geometric spectral inversion is to reconstruct $f$
 from $F.$\medskip
\np
\hbox{\vbox{\psfig{figure=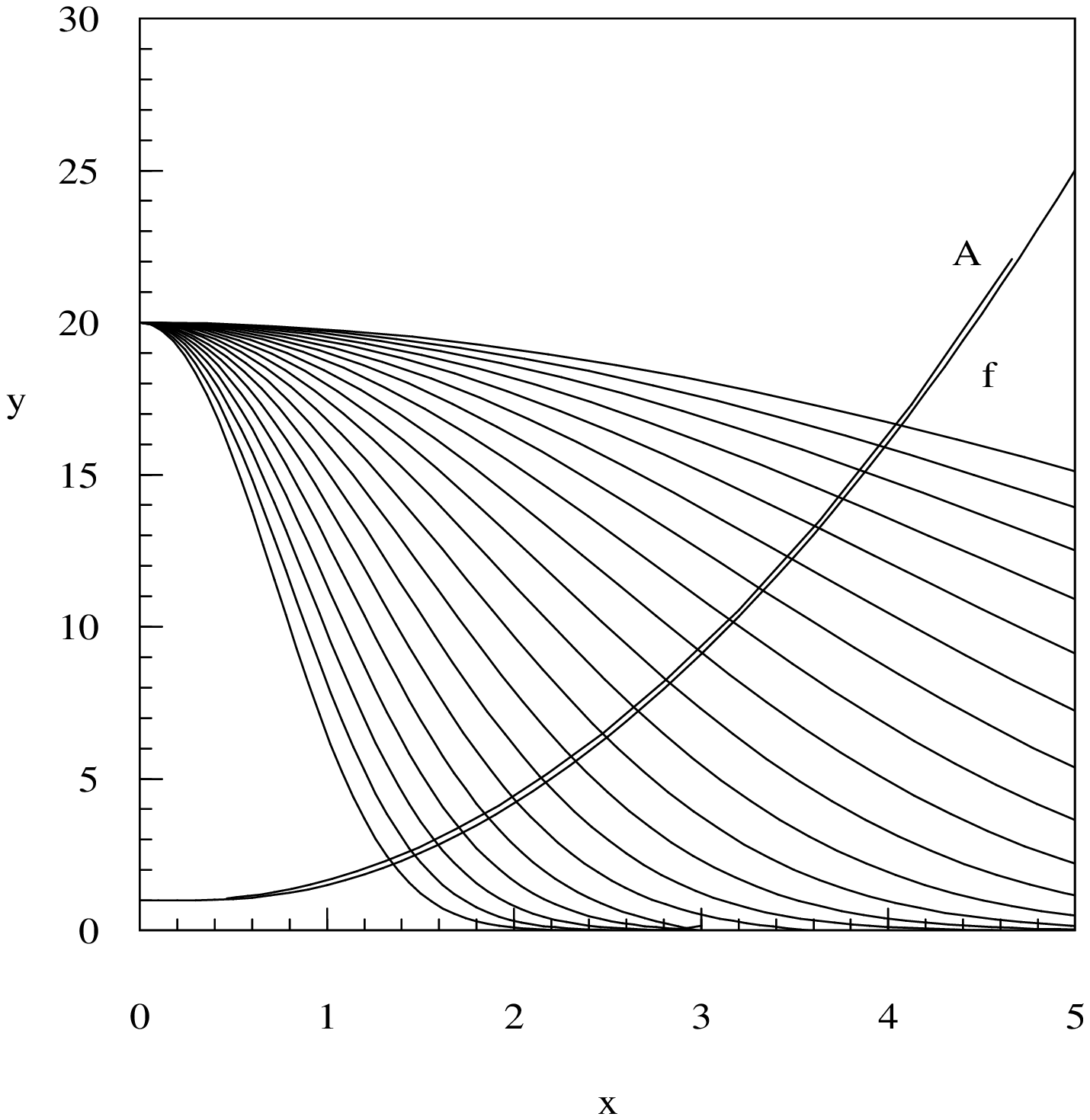,height=6in,width=6in,silent=}}}
\noindent{\bf Figure~(2)}~~The approximation $A$ obtained from the inversion
 inequality (2.5) is compared to the exact potential $f.$ The family of
 corresponding `exact' wave functions $\psi(x,v)$ satisfying $\psi(0,v) = 20$
 is also shown: the wave functions become monotonically concentrated towards
 zero as $v$ is increased from $v = 3\times {10}^{-4} \ {\rm to} \ 10.$ \medskip
% -------------------------------------------------------------------------
\hfil\vfil
\end